\documentclass[
 aip,
 pop,
amsfonts,
amsmath,
amssymb,
reprint,%
]{revtex4-1}

\usepackage[colorlinks,linkcolor=blue,urlcolor=blue,citecolor=blue]{hyperref}
\usepackage{textcomp}
\usepackage{amsmath,amstext,amsfonts,amssymb, bm}
\usepackage{graphicx}
\usepackage{gensymb}
\usepackage{float}
\usepackage{xcolor}
\usepackage{lineno}
\usepackage{multirow}
\usepackage{dcolumn}
\usepackage{etoolbox}

\makeatletter
\def\@email#1#2{%
 \endgroup
 \patchcmd{\titleblock@produce}
  {\frontmatter@RRAPformat}
  {\frontmatter@RRAPformat{\produce@RRAP{*#1\href{mailto:#2}{#2}}}\frontmatter@RRAPformat}
  {}{}
}%
\makeatother
\begin{document}

\title{On-the-fly machine learned force fields for the study of warm dense matter: application to diffusion and viscosity of CH}

\author{Shashikant Kumar}
 \affiliation{College of Engineering, Georgia Institute of Technology, Atlanta, GA 30332, USA}
\author{Xin Jing}
\affiliation{College of Engineering, Georgia Institute of Technology, Atlanta, GA 30332, USA}
\affiliation{College of Computing, Georgia Institute of Technology, Atlanta, Georgia 30332, USA}
\author{John E. Pask}
\affiliation{Physics Division, Lawrence Livermore National Laboratory, Livermore, CA 94550, USA}
\author{Phanish Suryanarayana}
\email[Email: ]{phanish.suryanarayana@ce.gatech.edu}
\affiliation{College of Engineering, Georgia Institute of Technology, Atlanta, GA 30332, USA}
\affiliation{College of Computing, Georgia Institute of Technology, Atlanta, Georgia 30332, USA}

\date{\today}

\begin{abstract}
We develop a framework for on-the-fly machine learned force field (MLFF) molecular dynamics  (MD) simulations of warm dense matter (WDM). In particular, we employ an MLFF scheme based on the kernel method and Bayesian linear regression, with the training data  generated from Kohn-Sham density functional theory (DFT)  using the Gauss Spectral Quadrature  method, within which we calculate energies,  atomic forces, and stresses.   We verify the accuracy of the formalism by comparing the predicted properties of  warm dense carbon  with recent Kohn-Sham DFT results in the literature. In so doing, we demonstrate that ab initio MD simulations of WDM can be accelerated by up to three orders of magnitude, while retaining ab initio accuracy. We apply this framework to calculate the diffusion coefficients and shear viscosity of CH at a density of 1 g/cm$^3$ and temperatures in the range of 75,000 to 750,000 K.  We find that the self- and inter-diffusion coefficients as well as the viscosity obey a power law with temperature, and that the diffusion coefficient results suggest a weak coupling between C and H in  CH. In addition, we find agreement within standard deviation with previous results for C and CH but disagreement for H, demonstrating the need for ab initio calculations as presented here.
\end{abstract}


\maketitle

\section{\label{sec:introduction}Introduction}

Warm dense matter (WDM) can be found in diverse physical settings, ranging from  giant planets and stars to inertial confinement fusion (ICF) and other high energy density (HED) experiments \cite{graziani2014frontiers}.  The accurate modeling of WDM is therefore crucial to the understanding and design of HED experiments like ICF, as well as to the understanding of the formation, nature, and evolution of planetary and stellar systems. However, such extreme conditions of temperature and pressure present significant challenges,  experimentally as well as theoretically, since both classical and quantum mechanical (degeneracy) effects contribute to the overall properties and behavior,  with the relative importance of each noticeably varying with the conditions, i.e., temperature and density, of interest. 

Kohn-Sham density functional theory (DFT) \cite{kohn1965self, hohenberg1964inhomogeneous} is a widely used method for studying materials systems from the first principles of quantum mechanics, without any empirical or \emph{ad hoc} parameters. However, Kohn-Sham calculations of WDM pose unique challenges, including the increase in the number of partially occupied states  with temperature, whereby the cubic scaling bottleneck of such methods manifests itself at smaller system sizes. This bottleneck becomes particularly restrictive in ab initio molecular dynamics (AIMD), where the Kohn-Sham equations may need to be solved hundreds of thousands of times to reach the timescales relevant to phenomena  of interest. This has motivated the development of formulations of Kohn-Sham DFT that are well suited to calculations at high temperature, including Spectral Quadrature (SQ) DFT \cite{bhattacharya2022accurate, suryanarayana2013spectral, suryanarayana2018sqdft, pratapa2016spectral}, stochastic DFT (SDFT) \cite{cytter2018stochastic, baer2013self}, mixed stochastic-deterministic DFT (MDFT) \cite{white2020fast}, and the density kernel based SQ method (SQ3) \cite{xu2022real}. In particular, the SQ method scales linearly with system size, has a prefactor that decreases rapidly with temperature, and has excellent parallel scaling, whereby it has found a number of applications in the study of WDM\cite{bethkenhagen2023properties, zhang2019equation, wu2021development}. 

In spite of significant advances, the relatively large computational cost associated with ab initio methods has prompted the development of a number of  approximations to Kohn-Sham DFT, including orbital-free molecular dynamics (OFMD) \cite{lamclegil2006}, extended first principles molecular dynamics (ext-FPMD) \cite{zhang2016extended, blanchet2021extended}, and spectral partitioned DFT (spDFT) \cite{sadigh2023spectral}.  However, though significantly reduced, these methods are still associated with significant computational cost, in the context of MD in particular. This limitation can be overcome by machine learned force field (MLFF) schemes \cite{unke2021machine, poltavsky2021machine, wu2023applications}, which have recently found use in  the study of WDM \cite{hinz2023development, liu2020structure, mahmoud2022predicting, kumar2023transferable, chen2023combining, tanaka2022development, nguyen2024extreme, willman2022machine}. However, such schemes generally require an extensive training dataset,  comprised of tens to hundreds of thousands of atomic configurations \cite{zhang2020warm, zeng2021ab,liu2021thermal, mahmoud2022predicting}, which is not only a computationally and labor intensive process, but generally  needs to be repeated for different conditions.  This limitation can be overcome by employing on-the-fly MLFF training during molecular dynamics (MD) simulations \cite{jinnouchi2019fly, jinnouchi2020fly, verdi2021thermal, liu2021alpha, Chen2023transferable, kumar2023kohn}. However, the efficiency and efficacy of such a scheme has not explored in the context of WDM heretofore.

In this work, we develop a framework for on-the-fly MLFF MD simulations. In particular, we employ an MLFF scheme based on the kernel method and Bayesian linear regression, with the Gauss SQ method used for generation of the Kohn-Sham  training data, which includes the energies, atomic forces, and stresses.  Through comparisons with recent Kohn-Sham results in the literature, we show that the framework is able to accelerate AIMD simulations of WDM by up to three orders of magnitude, while retaining ab initio accuracy. We apply this framework to calculate the diffusion coefficients and shear viscosity of C, H, and CH at a density of 1 g/cm$^3$ and temperatures in the range of 75,000 to 750,000 K, where we find agreement with previous results for C and CH but disagreement for H, demonstrating the need for ab initio calculations as presented here.

The remainder of this paper is organized as follows. In Section~\ref{sec:formulation}, we discuss the formulation and implementation of the framework for on-the-fly MLFF MD simulations of WDM.  In Section~\ref{Sec:Results}, we first verify its accuracy and performance, and then apply it the study of warm dense CH. Finally, we conclude in Section~\ref{sec:conclusion}. 

\section{\label{sec:formulation}Formulation and Implementation}
\begin{figure}[htbp]
\includegraphics[width=0.95\linewidth]{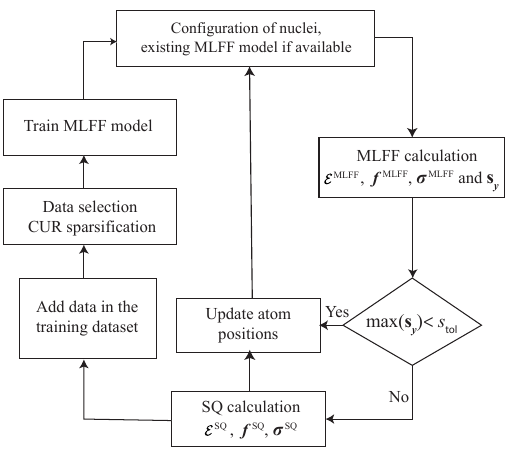}
        \caption{Outline of the SQ-MLFF framework.}
        \label{fig:flowchart}
\end{figure}

We have developed the framework outlined in Fig.~\ref{fig:flowchart}  for performing on-the-fly MD simulations of WDM, referred to as SQ-MLFF. In particular, we employ the Gauss variant of the SQ method for generation of the Kohn-Sham DFT data for training, which includes energies, atomic forces, and stresses, as described in  Section~\ref{Subsec:GaussSQ}; and an MLFF scheme based on the kernel  method and Bayesian linear regression for prediction of the quantities of interest, as described in Section~\ref{Subsec:MLFF}.  The quantities that form a part of both training and prediction are the energy ($\mathcal{E}$), atomic forces ($\boldsymbol{f}$), and stress tensor  ($\boldsymbol{\sigma}$), where the superscript, if present, denotes the method used for its calculation.

The MD simulation starts with a few SQ calculations that provide the initial training data. This is then  followed by predictions from the MLFF model in subsequent MD steps, except when the uncertainty in forces so computed, i.e., Bayesian error,  exceeds the threshold $s_{\rm tol}$, at which point an SQ calculation is again performed, data from which is added to the training dataset. The threshold $s_{\rm tol}$ is not fixed, but is set to the maximum value of the Bayesian error in forces for the MLFF based MD step subsequent to the training step. To avoid the cubic scaling bottleneck in training, a two-step data selection procedure is adopted, wherein only those atoms  whose Bayesian error in forces exceeds a predefined threshold are added to the training dataset, and then CUR \cite{mahoney2009cur} is performed on the resulting dataset for downsampling.


\subsection{Gauss Spectral Quadrature (SQ) method \label{Subsec:GaussSQ}}
The finite-temperature density matrix in Kohn-Sham DFT  takes the form:
\begin{align} \label{Eq:DensityMatrix}
\boldsymbol{D} = f(\boldsymbol{H}) = \left(1 + \exp \left( \frac{\boldsymbol{H} - \mu \boldsymbol{I}}{k_B T} \right) \right)^{-1} \,,
\end{align}
where $f$ is the Fermi-Dirac distribution, $\boldsymbol{H}$ is the Hamiltonian matrix that itself depends on part of the density matrix, $\boldsymbol{I}$ is the identity matrix, $k_B$ is the Boltzmann constant, $T$ is the electronic temperature, and $\mu$ is the chemical potential, obtained by enforcing the constraint on the number of electrons in the cell. Upon self-consistent solution, quantities of interest such as the free energy, Hellmann-Feynman atomic forces, and Hellmann-Feynman stress tensor can be computed from the density matrix \cite{suryanarayana2018sqdft, sharma2020real}.

In the Gauss SQ method \cite{bhattacharya2022accurate, suryanarayana2013spectral, Phanish2012}, the quantities of interest are written as bilinear forms or sums of bilinear forms, which are then approximated by Gauss quadrature rules. Consider for instance the electron density, which is the diagonal of the density matrix in a real-space finite-difference representation. Neglecting truncation, its value at the $n^{th}$ finite-difference node is approximated  as:
\begin{align} \label{Eq:Dens:SQ}
\boldsymbol{e}_n^{T} \boldsymbol{D} \boldsymbol{e}_n \approx   \sum_{k=1}^{N_q} w_{n,k}  f(\lambda_{n,k}) \,,
\end{align}
where $e_n$ denotes the standard basis vector, $N_q$ is the quadrature order, $\lambda_{n,k}$ are the quadrature nodes, and $w_{n,k}$ are the quadrature weights. In particular, the nodes are the eigenvalues,  and the weights are the   squares of the first elements of the  eigenvectors of the  Jacobi matrix $\boldsymbol{J}_n$ formed during the Lanczos iteration \cite{bhattacharya2022accurate, suryanarayana2013spectral, Phanish2012}.  Such a strategy is applicable to bilinear forms involving other functions of the Hamiltonian, such as those appearing in the band structure energy and electronic entropy. Indeed,  the nodes and weights are agnostic to the function being integrated, whereby they need to be determined only once. The Gauss SQ method has thus been used  to calculate  the density and the ground state energy in Kohn-Sham DFT. 

The Gauss SQ  formalism as described  is not amenable to the efficient calculation of the off-diagonal components of the density matrix, which are required for the calculation of the  Hellmann-Feynman atomic forces and stresses \cite{sharma2020real, pratapa2016spectral, suryanarayana2018sqdft}, quantities that are of particular importance in MD simulations. To overcome this limitation, we observe that the Lanczos iteration can be interpreted as performing the following decomposition of the Hamiltonian:
\begin{align}
\boldsymbol{H} \approx \boldsymbol{V}_{n} \boldsymbol{J}_n \boldsymbol{V}_{n} ^T \,,
\end{align}
where $ \boldsymbol{V}_{n}$ is a matrix of the orthonormal vectors generated during the Lanczos iteration, with the starting vector $\boldsymbol{e}_n$ forming the first column. It therefore follows that the $n^{th}$ column of the density matrix can be written as:
\begin{align} 
\boldsymbol{D} \boldsymbol{e}_n \approx  \boldsymbol{V}_{n} f(\boldsymbol{J}_n) \boldsymbol{V}_{n} ^T \boldsymbol{e}_n =   \boldsymbol{V}_{n} f(\boldsymbol{J}_n)  \boldsymbol{e}_1  \,.
\end{align}
Since all these quantities  are already available as part of the current Gauss SQ formulation, the calculation of the off-diagonal components of the density matrix, and therefore the atomic forces and stresses, incurs negligible additional cost. In particular, this formalism not only provides significant computational savings relative to the Clenshaw-Curtis SQ method \cite{bhattacharya2022accurate, suryanarayana2018sqdft, pratapa2016spectral, suryanarayana2013spectral, sharma2020real}, used previously for the calculation of the forces and stresses, but also noticeably increases the simplicity of the implementation and its use in production calculations.

In its complete form, the Gauss SQ method exploits the nearsightedness of matter \cite{prodan2005nearsightedness}, neglected here for simplicity of presentation.  In particular, within a real-space representation, the finite-temperature density matrix has exponential decay away from its diagonal  \cite{goedecker1998decay, ismail1999locality, benzi2013decay, suryanarayana2017nearsightedness}, a consequence of the locality of electronic interactions. This can be exploited to restrict the bilinear forms to be spatially localized, i.e., use of nodal Hamiltonians \cite{goedecker1998decay, suryanarayana2018sqdft, pratapa2016spectral} rather than the full Hamiltonian, which enables the Gauss SQ method to scale linearly with system size, with increasing efficiency at higher temperatures  as the density matrix becomes more localized and the Fermi operator becomes smoother \cite{suryanarayana2018sqdft, suryanarayana2017nearsightedness, pratapa2016spectral}. On increasing  quadrature order and localization radius,  convergence to exact cubic scaling diagonalization results is readily obtained \cite{sharma2020real, suryanarayana2018sqdft, pratapa2016spectral}. The SQ method also provides results corresponding to the infinite-crystal  without recourse to Brillouin zone integration or large supercells \cite{Phanish2012, suryanarayana2018sqdft, pratapa2016spectral}, a technique referred to as the infinite-cell method.


\subsection{Machine learned force field (MLFF) \label{Subsec:MLFF}}
The energy in the MLFF scheme takes the form \cite{bartok2010gaussian, kumar2023kohn}:
\begin{equation} 
    \mathcal{E}^{\rm MLFF} = \sum_{E} \sum_{i=1}^{N_A^E} \varepsilon_i^E  = \sum_E \sum_{i=1}^{N_A^E} \sum_{t=1}^{N_{T}^{E}} \tilde{w}_{t}^{E} k\left(\mathbf{x}_{i}^{E}, \tilde{\mathbf{x}}_t^{E} \right) \,,
    \label{Eq:EnergyDecompML}
\end{equation}
where $E$ denotes the chemical element that also serves as an index, $\varepsilon_i^E$ are the atomic energies, $\tilde{w}_{t}^{E}$ are the model weights, and $k\left(\mathbf{x}_{i}^{E}, \tilde{\mathbf{x}}_t^{E} \right)$ is a kernel that measures the similarity of the descriptor vectors $\mathbf{x}_{i}^{E}$ and $\tilde{\mathbf{x}}_t^{E}$, the former corresponding to the $N_A^E$ atoms in the current atomic configuration, and the latter corresponding to the $N_{T}^{E}$  atoms in the  training dataset. In this work, we employ the polynomial kernel and the Smooth Overlap of Atomic Positions (SOAP)  descriptors \cite{bartok2013representing}. Since the dependence of the  descriptors on the atomic positions is explicitly known, the corresponding atomic forces $\boldsymbol{f}^{\rm MLFF}$ and stress tensor $\boldsymbol{\sigma}^{\rm MLFF}$ are immediately   accessible,  detailed expressions for which can be found in the literature \cite{kumar2023kohn}.

The weights in the machine learned model are determined during training through Bayesian linear regression \cite{bishop2006pattern}:
\begin{equation} 
    \tilde{\boldsymbol{w}} = \beta(\alpha \boldsymbol{I} + \beta \tilde{\boldsymbol{K}}^{T} \tilde{\boldsymbol{K}})^{-1}  \tilde{\boldsymbol{K}}^{T} \tilde{\boldsymbol{y}} \,,
    \label{Eq:wtsML}
\end{equation}
where $\tilde{\boldsymbol{w}}$ is the vector of weights; $\alpha$ and $\beta$ are parameters that are determined by maximizing the evidence function\cite{bishop2006pattern}; $\tilde{\boldsymbol{y}}$ is the vector containing the quantities of interest, i.e., energy, atomic forces, and stresses --- suitably normalized by their mean and standard deviation values \cite{kumar2023kohn} --- for the atomic configurations in the training dataset; and $\tilde{\boldsymbol{K}}$ is the corresponding covariance matrix. 

The energy, atomic forces, and stresses for a given atomic configuration,  collected in the vector  $\mathbf{y}$, can then be predicted from the machine learned model using the relation:
\begin{equation} 
    \boldsymbol{y} = \boldsymbol{K} \tilde{\boldsymbol{w}} \,,
    \label{Eq:LinearSystemML}
\end{equation}
where $\mathbf{K}$ represents the covariance matrix associated with the given atomic configuration. The uncertainty in the values so predicted can be ascertained using the relation
\begin{equation} 
    \boldsymbol{s}^2_{\boldsymbol{y}} = {\rm diag}\left( \frac{1}{\beta} \boldsymbol{I} + \boldsymbol{K} (\alpha \boldsymbol{I} + \beta \tilde{\boldsymbol{K}}^{T} \tilde{\boldsymbol{K}})^{-1} \boldsymbol{K}^{T} \right)\,,
    \label{Eq:UQML}
\end{equation}
where $\mathbf{s}_{\mathbf{y}}$ is a vector with the  Bayesian error in the energy, atomic forces, and stresses as entries. 


\subsection{Ionic transport properties}
The SQ-MLFF framework is amenable to the calculation of ionic transport properties, i.e., diffusion coefficients, shear viscosity, and ionic thermal conductivity. In particular, these quantities can be determined from MD simulations using the Green-Kubo relations, wherein macroscopic dynamics properties are written as time-integrals of microscopic time correlation functions \cite{rapaport2004art}. Though the ionic thermal conductivity --- a quantity that is incompatible with the DFT formalism --- becomes accessible in  SQ-MLFF, it is not computed in the current work, since the electronic contribution is expected to be more dominant, particularly for the conditions of interest \cite{liu2021thermal}. 


The self-diffusion coefficient can be obtained by integrating the velocity autocorrelation function \cite{rapaport2004art}:
\begin{equation} 
    D_{\rm E} = \frac{1}{3N_{A}^{E}} \int_{0}^{\infty} dt \sum_{i = 1}^{N_{A}^E} \langle \bm{v}_i^E(0) \cdot  \bm{v}_i^E(t) \rangle\,, 
    \label{Eq:SelfDiffCoeff}
\end{equation}
where  $\bm{v}_i^E$ is the velocity vector of the $i$-th atom of element type $E$, and $\langle . \rangle$ represents the ensemble average. In the case of binary mixtures, the inter-diffusion coefficient can also be defined, which can be computed by integrating the diffusion velocity autocorrelation function\cite{GRABOWSKI2020100905,boercker1987interdiffusion}:
\begin{equation} 
   D_{\rm E_1 E_2}= \frac{1}{3(N_{A}^{E_1}+N_{A}^{E_2}) S_{cc}} \int_{0}^{\infty} dt \langle\bm{v}_d(0) \cdot  \bm{v}_d(t)\rangle \,, 
    \label{Eq:InterDiff}
\end{equation}
where $E_1$ and $E_2$ denote the element types, $\bm{v}_d$ is the diffusion velocity:
\begin{equation} 
   \bm{v}_d(t) = c_2 \sum_{i=1}^{N_A^{E_1}} \bm{v}_{i}^{E_1}(t) - c_1 \sum_{i=1}^{N_A^{E_2}} \bm{v}_i^{E_2}(t) \,, 
    \label{Eq:vdInterDiff}
\end{equation}
and $S_{cc}$ is the zero wavevector concentration structure factor, which  can be written in terms of the zero wavevector partial structure factor $S_{E_1E_1}$,  $S_{E_2 E_2}$, and $S_{E_1 E_2}$ as:
\begin{equation} 
   S_{cc} = c_{E_1} c_{E_2} [c_{E_2} S_{E_1E_1} + c_{E_1} S_{E_2 E_2} - 2 \sqrt{c_{E_1} c_{E_2}} S_{E_1 E_2}] \,, 
    \label{Eq:ScInterDiff}
\end{equation}
with $c_{E_1}$ and $c_{E_2}$ representing the molar fractions of elements $E_1$ and $E_2$, respectively.

The shear viscosity can be calculated by integrating the shear stress autocorrelation function, which can also be formulated as \cite{alfe1998first}: 
\begin{equation} 
    \eta = \frac{V}{5k_B T} \int_{0}^{\infty} dt \sum_{i=1}^5 \langle \sigma'_i(0) \cdot  \sigma'_i(t)\rangle\,, 
    \label{Eq:ViscCoeff}
\end{equation}
where $V$ is the volume of the cell, $T$ is the ionic temperature, and $\bm{\sigma}' = [ \sigma_{12}, \sigma_{23}, \sigma_{31}, (\sigma_{11}-\sigma_{22})/2, (\sigma_{22}-\sigma_{33})/2]$ is a vector containing the five independent components of the deviatoric stress tensor.


\section{\label{Sec:Results} Results and discussion}

We have developed a parallel implementation the SQ-MLFF formalism in the SPARC electronic structure code \cite{ghosh2017sparc1, ghosh2017sparc2, xu2021sparc, zhang2023version}. Choosing diffusion coefficient and shear viscosity as the target properties of interest, we first verify the accuracy and performance of SQ-MLFF for warm dense carbon (C) in Section~\ref{sec:accuracyperformance}, and then apply it to the study of a warm dense carbon hydrogen mixture (CH)  in Section~\ref{sec:CHresults}. The associated data can be found in Appendix~\ref{App:Data}. In all cases, we perform isokinetic ensemble (NVK) MD simulations with the Gaussian thermostat \cite{minary2003algorithms} for 500,000 steps, with the initial 5000 steps used for equilibration and the remaining used for production. In so doing, the statistical errors in the self-diffusion coefficient, inter-diffusion coefficient, and viscosity are reduced to within $4 \times 10^{-4}$ cm$^2$/s, $0.005$ cm$^2$/s, and 0.1 mPa~s, respectively. 

In the DFT calculations, we adopt the local density approximation (LDA) for the exchange-correlation \cite{kohn1965self, perdew1981self}, and optimized norm-conserving Vanderbilt (ONCV) pseudopotentials \cite{hamann2013optimized} with 1 and 6 electrons in valence for H and C, respectively, specifically constructed to be accurate for the conditions under consideration \cite{suryanarayana2023accuracy}. All DFT parameters, including the mesh size, and the Gauss SQ parameters, namely, truncation radius  and quadrature order, are chosen such that the forces and stresses are converged to within 1\% and 2\%, respectively. This translates to  the discretization error in the diffusion coefficients and viscosity being within 0.5\% and 1\%, respectively. Note that the error further decreases with higher grid resolution, truncation radius, and quadrature order.  In the MLFF calculations, the hyperparameters are chosen to be the same as that found optimal in previous work  \cite{kumar2023kohn}. Indeed, the performance of the MLFF scheme has been found to relatively insensitive to their choice, even for the extreme conditions considered  here. 

\subsection{\label{sec:accuracyperformance}Accuracy and Performance}

We consider C at a density of 10 g/cm$^3$ and temperatures in the range of 100,000 to 2,000,000 K, specifically, 100,000, 200,000, 500,000, 750,000, 1,000,000, and 2,000,000 K. In the MD simulations, we consider system sizes of 200 and 64 atoms for the lowest and highest temperatures, respectively, and interpolate linearly for temperatures in between. MD time steps ranged from 0.09 and 0.019 fs, depending on temperature.

\begin{figure*}[htbp]
\includegraphics[width=0.66\linewidth]{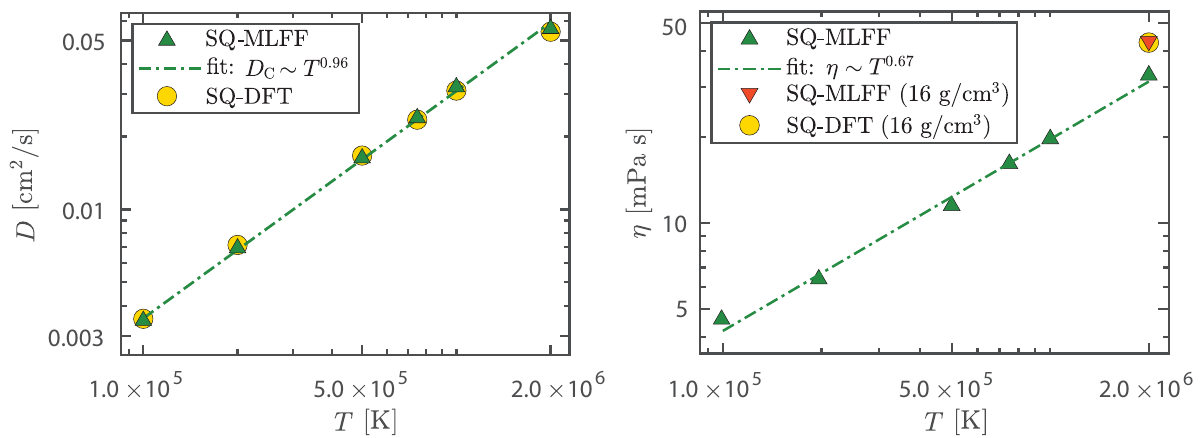}
        \caption{Variation of the diffusion coefficient and viscosity with temperature for C at a density of 10 g/cm$^3$, as calculated using the SQ-MLFF and SQ-DFT methods. The SQ-DFT data has been taken from Ref.~[\onlinecite{bethkenhagen2023properties}].}
        \label{fig:validation}
\end{figure*}

In Fig.~\ref{fig:validation}, we present the diffusion coefficient and viscosity computed using SQ-MLFF, and compare it with recent Kohn-Sham DFT results in the literature \cite{bethkenhagen2023properties}, obtained using the SQ method. We observe that there is very good agreement between the SQ-MLFF and SQ-DFT results, with a maximum difference of 0.002 cm$^2$/s in the diffusion coefficient. The viscosity, which was only computed  at a density of 16 g/cm$^3$ and 2,000,000 K in the cited reference, is also in very good agreement when the same conditions are simulated here, with a difference of only 0.4 mPa s, i.e., 43.2 mPa s from SQ-MLFF vs. 42.8 mPa s from SQ-DFT.  We also observe that both the diffusion coefficient and viscosity demonstrate a power law behavior with temperature, in agreement with previous results for dense plasmas \cite{sjostrom2015ionic, ticknor2015transport, hou2021ionic}. In particular, we obtain a exponent of 0.96 for the diffusion coefficient, which is in very good agreement with the exponent of 0.95 obtained from SQ-DFT \cite{bethkenhagen2023properties}. These results demonstrate the accuracy of SQ-MLFF for the calculation of ionic transport properties of WDM. Though not the focus of the work, the accuracy of SQ-MLFF extends to equations of state calculations, with the pressure computed in the above simulations being in agreement with SQ-DFT results to within 0.6\% (Appendix~\ref{App:Data}). 


\begin{table}[htbp]
    \centering
    \begin{tabular}{|c|c|c|c|c|c|}
        \hline
        \multirow{2}{*}{$T$ [K]} & \multicolumn{2}{c|}{\# MD steps}   & \multicolumn{2}{c|}{Time [CPU s]}  &  \multirow{2}{*}{Speedup} \\
        \cline{2-5}
              &  MLFF    & SQ &   MLFF      & SQ &    \\
        \hline
        100,000   & 499,694 & 306  & 17 & 7$\times 10^5$  & 1339 \\
        200,000   & 499,720 & 280  & 14 & 1$\times 10^5$  & 1246 \\
        500,000   & 499,724 & 276 & 10 & 4$\times 10^4$  & 1134 \\
        750,000   & 499,776 & 224 & 10 & 2$\times 10^4$  & 979 \\
        1,000,000 & 499,807 & 193 & 7 & 1$\times 10^4$  & 940 \\
        2,000,000 & 499,890 & 110 & 7 & 1$\times 10^4$  & 976\\
        \hline
    \end{tabular}
    \caption{Variation of SQ-MLFF performance with temperature for C at a density of 10 g/cm$^3$. The timings are per MD step, as averaged over the entire simulation. The speedup represents the ratio of the time taken for the MD simulation by SQ-DFT and SQ-MLFF, with the time taken for SQ-DFT estimated via extrapolation from 100 steps.}
    \label{Tab:MLFFPerformanceTable}
\end{table}

\begin{figure}[htbp]
\includegraphics[width=0.66\linewidth]{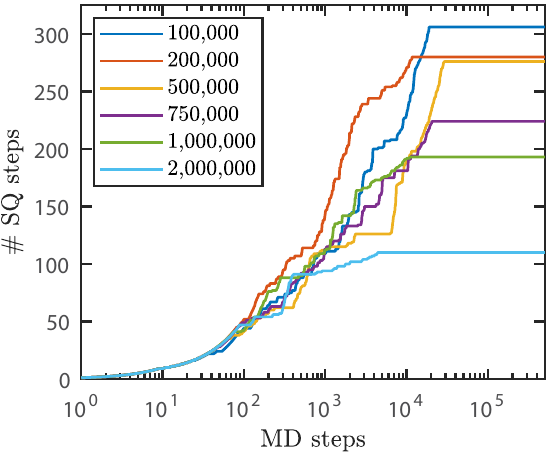}
        \caption{Cumulative number of DFT  steps during the SQ-MLFF MD simulation  for C at 10 g/cm$^3$, at different temperatures.}
        \label{fig:SQsteps_validation}
\end{figure}

In Table~\ref{Tab:MLFFPerformanceTable}, we present the performance of SQ-MLFF for the  MD simulations described above. We observe that the number of DFT steps, which constitutes of only $\sim$0.06\% and $\sim$0.02\% of the total number of MD steps for the lowest and highest temperatures, respectively,  decreases with temperature. This can likely be attributed to the decrease in quantum mechanical effects with temperature, making it more amenable to MLFF development. The occurrence of these DFT steps during the MD simulation can be found in Fig.~\ref{fig:SQsteps_validation}. We observe that most of the DFT calculations occur towards the beginning of the MD, with decreasing frequency as the simulation progresses. This is similar to the observations made for ambient conditions \cite{jinnouchi2019fly, jinnouchi2020fly, kumar2023kohn}, with one noticeable difference being that in the current simulations, a significantly larger number of DFT steps are performed even after a few thousand MD steps, likely due to the increased movement of atoms causing \emph{new} configurations to be  encountered later in the MD simulation. Even though there are so few DFT steps, they still constitute $\sim$90-95\% of the total CPU time. In terms of efficiency, it is estimated that there is speedup of  three orders of magnitude and more by SQ-MLFF relative to SQ-DFT in both CPU and wall time. Though the SQ formalism scales to many tens of thousands of processors and beyond \cite{gavini2023roadmap, bhattacharya2022accurate}, the current MLFF implementation scales only  up to the number of processors equal to the number of atoms. Therefore, the percentage of total CPU time taken by DFT and the speedup in wall time achieved by MLFF will reduce as the number of processors increases beyond the number of atoms in the system. Indeed, the MLFF code can be parallelized to scale to an order of magnitude larger number of processors, ensuring that three orders of magnitude speedup is also achieved in the wall time for such simulations. The speedup of SQ-MLFF increases with the number of MD steps, therefore the numbers for simulations targeting the diffusion coefficient alone will be be significantly smaller, e.g., the speedup to achieve a statistical error of  1\%, which requires around $\sim$25,000 steps, is estimated to be around two orders of magnitude. 


\subsection{\label{sec:CHresults}Application:  warm dense CH}

We consider CH at a density of 1 g/cm$^3$ and temperatures in the range 75,000 to 750,000 K, specifically, 75,000, 100,000, 200,000, 500,000, and 750,000 K. In the MD simulations, we consider system sizes of 216 and 108 atoms for the lowest and highest temperatures, respectively, and again interpolate linearly for temperatures in between. MD time steps ranged from 0.04 and 0.004 fs, depending on temperature. The Coulomb coupling parameter ($\Gamma$) and electron degeneracy parameter ($\Theta$) so calculated from the DFT steps for the conditions considered here range from 0.92 to 1.60 and 0.90 to 4.90, respectively (Appendix~\ref{App:Data}),  which falls within the WDM regime \cite{stanek2024review}. The computed pressure can also be found in Appendix~\ref{App:Data}. 

\begin{figure*}[htbp]
\includegraphics[width=0.9\linewidth]{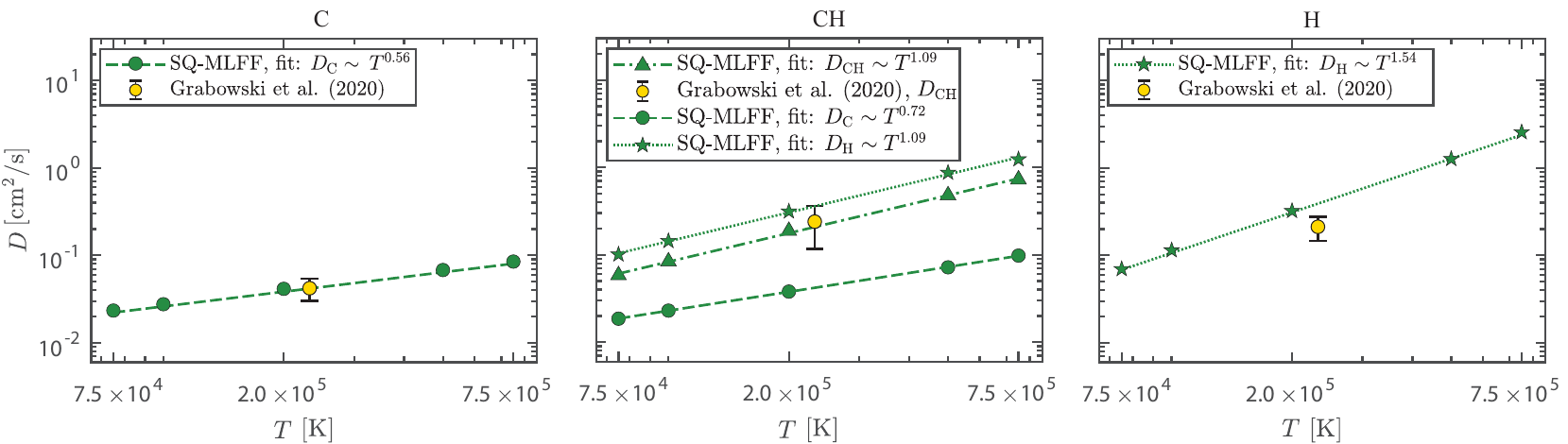}
        \caption{Variation of the diffusion coefficients with temperature for CH, C, and H, all at a density of 1 g/cm$^3$, as calculated using the SQ-MLFF method. The markers and error bars for the results of Grabowski et al. [\onlinecite{GRABOWSKI2020100905}] represent the mean and standard deviation, respectively, over the various methods compared in that work, which includes the average atom method and orbital-free DFT.}
        \label{fig:D_CH}
\end{figure*}

In Fig.~\ref{fig:D_CH}, we present the variation of the diffusion coefficients for CH --- self-diffusion coefficients for C  and H, and inter-diffusion coefficient for CH --- with temperature. We also compare these results with those obtained for C and H at the same conditions, again computed using SQ-MLFF.  We observe that the inter-diffusion coefficient has a power law behavior with temperature of $D_{\rm CH} \sim T^{1.09}$, with values that are closer to the self-diffusion coefficient of H than C in CH. Such a power law behavior is also observed for C and H, consistent with observations for other materials in the literature \cite{ticknor2015transport, hou2021ionic, sjostrom2015ionic, ticknor2016transport, white2019multicomponent}. In addition, we find that at each temperature, the value of the inter-diffusion coefficient is relatively close to the molar fraction weighted average of the self-diffusion coefficients of the individual species, i.e.,  $D_{\rm CH} \sim \frac{1}{2}(D_{\rm C} + D_{\rm H})$, with a maximum difference of 0.06 cm$^2$/s, suggesting  a relatively weak coupling between C and H in CH. This is substantiated by the self-diffusion coefficients of C and H being relatively close to their corresponding values in CH. We also observe that at 232,120 K, the value of $D_{\rm CH}$ as obtained from the power law fit is 0.21 cm$^2$/s, which is in relatively good agreement with recent work \cite{GRABOWSKI2020100905}, where the mean value between different methods, including the average atom method and orbital-free DFT, is 0.24 cm$^2$/s, with a relatively large standard deviation of 0.12 cm$^2$/s. Though this is also true for C, there is a significant difference in the predictions for H, suggesting that  higher fidelity is required for  simulations of H in the WDM regime.

\begin{figure*}[htbp]
\includegraphics[width=0.9\linewidth]{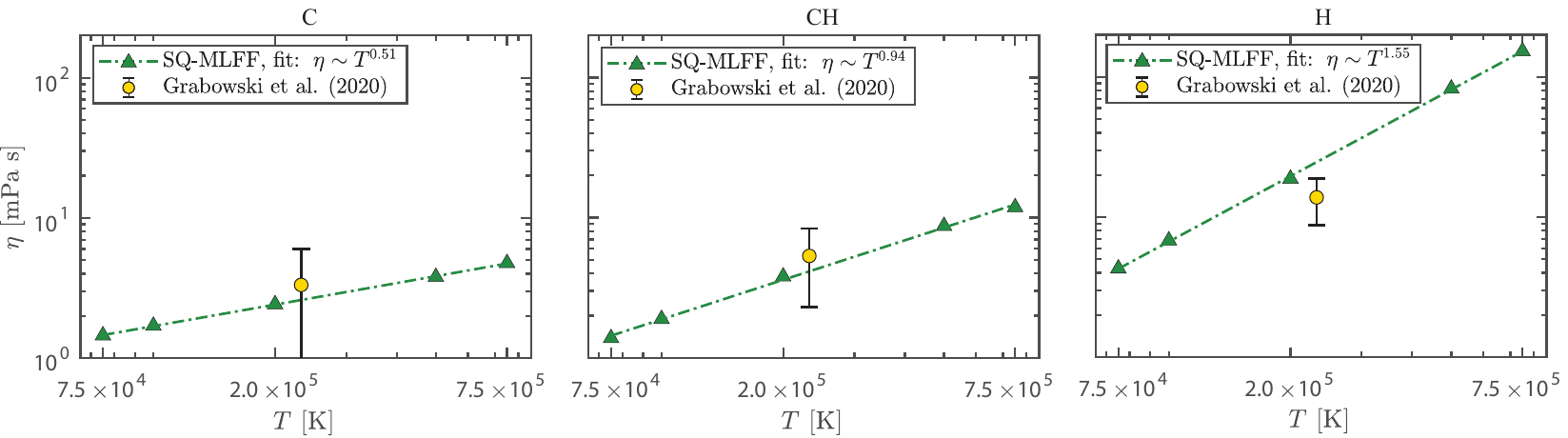}
        \caption{Variation of the viscosity with temperature for CH, C, and H, all at a density of 1 g/cm$^3$, as calculated using the SQ-MLFF method. The markers and error bars for the results of  Grabowski et al. [\onlinecite{GRABOWSKI2020100905}] represent the mean and standard deviation, respectively, over the various methods compared in that work, which includes the average atom method and orbital-free DFT.}
        \label{fig:viscosity_CH}
\end{figure*}

In Fig.~\ref{fig:viscosity_CH}, we present the variation of the viscosity with temperature for CH. We also compare these results with the viscosity of C and H at these conditions, again computed using SQ-MLFF. We observe that the viscosity has a power law behavior with temperature of $\eta \sim T^{0.94}$, with values that are closer to the viscosity of C. Such a power law behavior has been observed for other materials in the literature \cite{hou2021ionic, sjostrom2015ionic}. We also observe that at 232,120 K, the value of the viscosity as obtained from the power law fit is 4.14 mPa s, which is in relatively good agreement with recent work \cite{GRABOWSKI2020100905}, where the mean value between different methods, including the average atom method and orbital-free DFT, is 5.33 mPa s, with a relatively large standard deviation of 3.03 mPa s. Though this is also true for C, there is again a significant difference for H, further substantiating the need for higher fidelity in the simulations of H in the WDM regime.


\section{\label{sec:conclusion}Concluding remarks}

In this work, we have developed a framework for on-the-fly MLFF MD simulations of WDM, wherein we have employed an MLFF scheme based on the kernel method, and Bayesian linear regression, with the required training data   generated from Kohn-Sham DFT  using the Gauss SQ  method, within which we calculate energies, atomic forces, and stresses. Choosing warm dense carbon as a representative example, we  verified the accuracy of the formalism through comparisons with recent Kohn-Sham DFT results in the literature. In so doing, we have demonstrated that AIMD simulations of WDM can be accelerated up to three orders of magnitude while retaining ab initio accuracy. We have applied this framework to calculate the diffusion coefficients and shear viscosity of CH at a density of 1 g/cm$^3$ and temperatures in the range of 75,000 to 750,000 K. We found that the self- and inter-diffusion coefficients as well as the viscosity obey a power law with temperature, and that the diffusion coefficients suggest that the coupling between C and H is relatively weak in  CH. In addition, we found agreement within standard deviation with previous results for C and CH but disagreement for H, demonstrating the need for ab initio calculations as presented here. Overall, the proposed formalism promises to  enable simulations of WDM with ab initio accuracy at a small fraction of the original cost. 

The inclusion of internal energy within the machine learned model will help accelerate equation of state calculations such as the Hugoniot, making it a worthy subject for future research. Extending the current MLFF implementation to enable efficient scaling on large scale supercomputers, with GPU acceleration of the key computational kernels \cite{sharma2023gpu}, will further reduce the wall time for large MD simulations, making it another worthy subject for future research. 


\begin{acknowledgments}
The authors gratefully acknowledge support from grant DE-NA0004128 funded by the U.S. Department of Energy (DOE), National Nuclear Security Administration (NNSA).  J.E.P gratefully acknowledges support from the U.S. DOE, NNSA: Advanced Simulation and Computing (ASC) Program at Lawrence Livermore National Laboratory (LLNL). This work was performed in part under the auspices of the U.S. DOE  by LLNL under Contract DE-AC52-07NA27344. This research was also supported by the supercomputing infrastructure provided by Partnership for an Advanced Computing Environment (PACE) through its Hive (U.S. National Science Foundation through grant MRI-1828187) and Phoenix clusters at Georgia Institute of Technology, Atlanta, Georgia.
\end{acknowledgments}


\appendix
\section{\label{App:Data} SQ-MLFF data}
The SQ-MLFF data  for the C, H, and CH systems studied in this work is summarized in Table~\ref{Tab:Data}, where in addition to the diffusion coefficients and the viscosity, which are the focus of the current work, we also report the average ionization of ions, Coulomb coupling parameter, electron degeneracy parameter, and the total pressure. 

\begin{table*}[htbp]
    \centering
    \begin{tabular}{|c|c|c|c|c|c|c|c|c|c|c|}
        \hline
        \multirow{2}{*}{System} & \multirow{2}{*}{$\rho$ [g/cm$^3$]} & \multirow{2}{*}{$T$ [K]}  & \multirow{2}{*}{$Z^*$} & \multirow{2}{*}{$\Gamma$} & \multirow{2}{*}{$\Theta$} & \multirow{2}{*}{$P$ [TPa]} & \multicolumn{3}{c|}{$D$ [cm$^2$/s]} & \multirow{2}{*}{$\eta$ [mPa s]}  \\
        \cline{8-10}
            &   &   &  &  & &   & $D_{\rm C}$ & $D_{\rm CH}$ & $D_{\rm H}$ &   \\
        \hline
        \multirow{6}{*}{C} & 10 & 100,000  & 0.67 & 0.96 & 0.49 & 4.32 & 0.004 $\pm$ 0.00002 & - & - & 4.5  $\pm$ 0.04\\
                            & 10 & 200,000  & 1.22 & 1.59 & 0.66 & 6.92 & 0.007 $\pm$ 0.00002 & - & - & 6.3  $\pm$ 0.04 \\  
                           & 10 & 500,000   & 2.26 & 2.19 & 1.09 & 15.75 & 0.016 $\pm$ 0.00005 & - & - & 11.5  $\pm$ 0.04 \\  
                           & 10 & 750,000  & 2.77 & 2.19 & 1.42  & 23.86 & 0.024 $\pm$ 0.00006 & - & - & 16.2  $\pm$ 0.05 \\
                           & 10 & 1,000,000  & 3.18 & 2.16 & 1.73  & 32.96 & 0.032 $\pm$ 0.00009 & - & - & 19.7  $\pm$ 0.06\\
                           & 10 & 2,000,000  & 4.43 & 2.10 & 2.78 & 77.11 & 0.056 $\pm$ 0.00020   & - & - & 32.9  $\pm$ 0.06 \\
        \hline
        \multirow{6}{*}{CH} & 1 & 75,000  & 0.95 & 1.47 & 0.90 & 0.21 & 0.019 $\pm$ 0.00008 & 0.058  $\pm$ 0.003& 0.101 $\pm$ 0.0002 & 1.4 $\pm$ 0.02 \\
                           & 1 & 100,000  & 1.13 & 1.56 & 1.07 & 0.27 & 0.023 $\pm$ 0.00008 & 0.084  $\pm$ 0.003& 0.144  $\pm$ 0.0002 & 1.9   $\pm$ 0.02\\
                           & 1 & 200,000  & 1.62 & 1.60 & 1.69 & 0.66 & 0.038 $\pm$ 0.00009  & 0.188  $\pm$ 0.004& 0.314  $\pm$ 0.0003 & 3.8  $\pm$ 0.03 \\
                           & 1 & 500,000  & 2.15 & 1.13 & 3.50 & 1.90 & 0.072 $\pm$ 0.00010  & 0.482  $\pm$ 0.005 & 0.872  $\pm$ 0.0003 & 8.7  $\pm$ 0.03 \\
                           & 1 & 750,000   & 2.38 & 0.92 & 4.90 & 3.12 & 0.098 $\pm$ 0.00010  & 0.728  $\pm$ 0.005& 1.243  $\pm$ 0.0004 & 11.9  $\pm$ 0.04\\
        \hline
        \multirow{6}{*}{C} & 1 & 75,000  & 1.60 & 3.39 & 0.96 & 0.11 & 0.022 $\pm$ 0.00008 & - & -& 1.5  $\pm$ 0.02\\
                          & 1 & 100,000  & 1.88 & 3.51 & 1.15 & 0.15 & 0.026 $\pm$ 0.00008 & - & - & 1.7  $\pm$ 0.02 \\
                          & 1 & 200,000   & 2.52 & 3.15 & 1.89 & 0.39 & 0.039 $\pm$ 0.00010 & - & - & 2.4  $\pm$ 0.02\\
                          & 1 & 500,000  & 3.39 & 2.28 & 3.88 & 1.41 & 0.064 $\pm$ 0.00010 & - & - & 3.8  $\pm$ 0.03 \\
                          & 1 & 750,000  & 3.83 & 1.94 & 5.37 & 2.31 & 0.080 $\pm$ 0.00020 & - & - & 4.8  $\pm$ 0.03\\   
        \hline
        \multirow{6}{*}{H} & 1 & 75,000  & 0.26 &0.20 & 0.62 & 1.12 & - & - & 0.067 $\pm$ 0.00009 & 4.3  $\pm$ 0.03  \\
                           & 1 & 100,000  & 0.29 & 0.19 & 0.77 & 1.45 & - & - & 0.112 $\pm$ 0.00010& 6.8  $\pm$ 0.05\\
                           & 1 & 200,000  & 0.48 & 0.26 & 1.10 & 2.93 & - & - & 0.311 $\pm$ 0.00020& 18.8  $\pm$ 0.04 \\
                           & 1 & 500,000  & 0.74 & 0.25 & 2.05 & 7.67 & - & - & 1.223 $\pm$ 0.00030& 82.6  $\pm$ 0.09 \\
                           & 1 & 750,000  & 0.76 & 0.17 & 3.03 & 11.76 & - & - & 2.474 $\pm$ 0.00040 & 152.8  $\pm$ 0.10 \\
        \hline 
    \end{tabular}
    \caption{Average ionization of ions ($Z^*$), Coulomb coupling parameter ($\Gamma$), electron degeneracy parameter ($\Theta$), total pressure ($P$), diffusion coefficients ($D$), and viscosity ($\eta$) for the C, H, and CH systems studied in this work, as determined using the SQ-MLFF method. The statistical errors in pressures are within 0.02\%.}
     \label{Tab:Data}
\end{table*}

\section*{Data Availability Statement}
The data that support the findings of this study are available within the article and from the corresponding author upon reasonable request.

\section*{Author declarations}
The authors have no conflicts to disclose.

%


\end{document}